\documentclass[twocolumn,english,floatfix,aps,prd,showpacs]{revtex4-1}
\usepackage{inputenc}
\usepackage{color}
\usepackage{babel}
\usepackage{amsmath}
\usepackage{amssymb}
\usepackage{wasysym}
\usepackage{graphicx}
\usepackage{enumerate}
\usepackage{subfig}
\usepackage{caption}
\usepackage{ragged2e}
\DeclareCaptionJustification{justified}{\justifying}
\newenvironment{rcases}
  {\left.\begin{aligned}}
  {\end{aligned}\right\rbrace}

\begin{document}

\title{Casimir force between $\delta-\delta^{\prime}$ mirrors transparent
at high frequencies}

\author{Alessandra N. Braga}
\email{alessandrabg@ufpa.br}
\affiliation{Faculdade de F\'{i}sica, Universidade Federal do Par\'{a}, Bel\'{e}m, Par\'{a},
Brazil}

\author{Jeferson Danilo L. Silva}
\email{jeferson.silva@icen.ufpa.br}
\affiliation{Faculdade de F\'{i}sica, Universidade Federal do Par\'{a}, Bel\'{e}m, Par\'{a},
Brazil}

\author{Danilo T. Alves}
\email{danilo@ufpa.br}
\affiliation{Faculdade de F\'{i}sica, Universidade Federal do Par\'{a}, Bel\'{e}m, Par\'{a},
Brazil}

\date{\today}

\begin{abstract}
We investigate, in the context of a real massless scalar field
in $1+1$ dimensions, models of partially reflecting mirrors simulated by Dirac $\delta-\delta^{\prime}$
point interactions. 
In the literature, these models do not exhibit 
full transparency at high frequencies. 
In order to provide a more realistic feature for these models,
we propose a modified $\delta-\delta^{\prime}$ point interaction that enables full transparency in the limit of high frequencies. 
Taking this modified $\delta-\delta^{\prime}$ model into account, we investigate the Casimir force, comparing our results with
those found in the literature.
\end{abstract}


\maketitle

\section{Introduction} \label{introduction}

In 1948, Casimir predicted an attractive force exerted by the vacuum
fluctuations of the electromagnetic field between two neutral, parallel
and perfectly conducting plates \cite{Casimir-1948}. This phenomenon
is denominated Casimir effect and it was observed experimentally \cite{review}.
The problem of the electromagnetic
field in the presence of a perfectly conducting plate can be divided
essentially in two boundary condition (BC) problems: the vector potentials
representing the transverse electric and transverse magnetic polarizations
are associated, respectively, to the well-known Dirichlet and Neumann
conditions. Recently, these conditions have also been considered in
the literature related to the Casimir effect in the context of scalar
fields \cite{neumann-casimir-estatico}. In this case, the Casimir
force is attractive if both plates impose the Dirichlet or Neumann
BCs to the field, but it is repulsive if one mirror imposes the Dirichlet
and the other imposes the Neumann conditions \cite{hhh}. 
A continuous interpolation of the Dirichlet and Neumann conditions is accomplished
with the aid of the Robin BC, and the Casimir force between two parallel
plates imposing the Robin BC to a scalar field can be repulsive, attractive
or null \cite{romeo-saharian-2002}. A generalized
Robin BC, which includes a term with second-order time derivative
of the field, has been recently considered in the investigation of
the Casimir effect \cite{Fosco-2013}. 
The Robin BC also appears, for instance,
in classical mechanics in the problem of a string vibrating with small
amplitudes when one of its extremities is tied to a massless ring (which
slides with no friction along a vertical rod) coupled to a spring
\cite{Fosco-2013,Mintz-2006}. 

Dirichlet, Neumann and Robin
BCs are related to perfect mirrors. On the other hand, real mirrors
are naturally transparent at high frequencies \cite{Casimir-1948,Moore-1970}.
A way to model partially reflecting mirrors is via Dirac $\delta$ potentials \cite{Barton-I-II-1995,Munoz-2013}.
This idea has been used
in the investigation of the Casimir effect and dynamical Casimir effect
\cite{Scardicchio-2005,Barton-I-II-1995,Fosco-2009,Parashar-2012,Barton-2004,Bordag-2006,Bordag-2004,Fosco-2012,Milton-2013,Munoz-2013,Castaneda-2015,JD-2016,Fosco-2013}.
Specifically, for a real massless scalar field $\Phi$ in $1+1$ dimensions, the mirror,
located at $x=0$, can be modeled by the Lagrangian \cite{Barton-I-II-1995,Munoz-2013}
(hereafter $c=\hbar=1$), 
\begin{equation}
{\cal L}={\cal L}_{0}-\mu\delta(x)\Phi^{2}(t,x),
\label{lagrangeana-U}
\end{equation}
where ${\cal L}_{0}=(1/2)[(\partial_{t}\Phi)^{2}-(\partial_{x}\Phi)^{2}]$,
and $\mu\geq0$ is the coupling constant between the field and the
mirror. The reflection and transmission coefficients are given, respectively,
by \cite{Barton-I-II-1995} 
\begin{equation}
r_{\pm}(\omega)=-\frac{i\mu}{\omega+i\mu},\quad s_{\pm}(\omega)=\frac{\omega}{\omega+i\mu},
\label{r-s-delta}
\end{equation}
where the labels ``$+$'' and ``$-$'' indicate the scattering
to the right and to the left of the mirror, respectively. Note that,
in this case, $s_{+}(\omega)=s_{-}(\omega)$ and $r_{+}(\omega)=r_{-}(\omega)$.
From Eq. (\ref{r-s-delta}), one can see that the parameter $\mu$
controls the transparency of the mirror. In the limit $\mu\rightarrow\infty$, the mirror becomes perfectly
reflecting, $|r_{\pm}(\omega)|=1$, and the $\delta$ mirror
imposes the Dirichlet BC to the field in both sides, namely
\begin{equation}
\Phi_{+}(t,0^{+})=0\quad\text{and}\quad\Phi_{-}(t,0^{-})=0,\label{Dirichlet}
\end{equation}
where $\Phi_{+}(t,x)$ and $\Phi_{-}(t,x)$ represent the field in
the right and left sides of the mirror, respectively. Note that the
$\delta$ mirror is naturally transparent at high frequencies,
\begin{equation}
\lim_{\omega\rightarrow\infty}s_{\pm}(\omega)=1.\label{barton-transparencia}
\end{equation}

The use of $\delta-\delta^{\prime}$ potentials in the investigation of the Casimir effect was also considered in the literature (see Ref. \cite{Castaneda-2015}). A single $\delta-\delta^{\prime}$ mirror is described by the Lagrangian density 
\begin{equation}
{\cal L}={\cal L}_{0}-[\mu\delta(x)+\lambda\delta^{\prime}(x)]\Phi^{2}(t,x),
\label{U-Munoz-0}
\end{equation}
where $\mu\geq0$ and $\lambda\in\mathbb{R}$ are the coupling constants. The reflection and transmission coefficients are given, respectively, by \cite{Castaneda-2015} 
\begin{equation}
r_{\mathrm{\pm}}(\omega)=\frac{\pm2\omega\lambda-i\mu}{\omega(1+\lambda^{2})+i\mu},\enskip s_{\pm}(\omega)=\frac{\omega(1-\lambda^{2})}{\omega(1+\lambda^{2})+i\mu}.\label{eq:rr-1-2}
\end{equation}
In this case, differently from the pure $\delta$ one [Eq. (\ref{lagrangeana-U})], $r_{+}(\omega)\neq r_{-}(\omega)$. Moreover, for $\lambda=1$, we get a perfectly reflecting mirror imposing the BCs 
\begin{eqnarray}
\mu\Phi_{+}(t,0^{+})+2\partial_{x}\Phi_{+}(t,0^{+}) & = & 0,\label{eq:RobinBC}\\
\qquad\Phi_{-}(t,0^{-}) & = & 0,\label{dirichlet}
\end{eqnarray}
which are identified as the Robin and Dirichlet BCs, respectively. 
Furthermore, another remarkable difference between $\delta$ and $\delta-\delta^{\prime}$
mirrors is that the latter is not naturally transparent at high frequencies, namely
\begin{equation}
\lim_{\omega\rightarrow\infty}s_{\mathrm{\pm}}(\omega)=\frac{1-\lambda^{2}}{1+\lambda^{2}}<1,\quad(\lambda\neq0).\label{eq:hhh}
\end{equation}
However, for sufficiently high frequencies, a real mirror does not act as an obstacle to the field \cite{Casimir-1948, Moore-1970,Jaekel-Reynaud-1991}, i.e. $\lim_{\omega\rightarrow\infty}|s_{\pm}(\omega)|=1$, and models with perfectly reflecting mirrors can lead to unphysical predictions (see, for instance, Refs. \cite{Haro-Elizalde-PRL-2006,Jaekel-Reynaud-1992-PLA}).
Moreover, for $\mu=0$ the scattering coefficients (\ref{eq:rr-1-2}) are independent of the frequency. These behaviors suggest that the model (\ref{U-Munoz-0}), which leads to (\ref{eq:hhh}), lacks adjustments to become more realistic.

In the present paper, in order to provide more realistic features for $\delta-\delta^{\prime}$ models,
we propose a Lagrangian density that describes $\delta-\delta^{\prime}$ mirrors 
with full transparency in the limit of high frequencies. 
Specifically, we do this in the context of a real massless scalar field in $1+1$ dimensions.
Taking this modified model into account, we investigate the Casimir force between two $\delta-\delta^{\prime}$ mirrors transparent at high frequencies. As we shall discuss, the 
transparency at high frequencies prevents spurious contributions, coming from the high-frequency modes, to be computed in the Casimir force, specially in the case of small distances between the mirrors.

The paper is organized as follows. In Sec. \ref{general}, we outline the essential aspects of the scattering matrix for a cavity with generic scattering coefficients, and the Casimir force \cite{Jaekel-Reynaud-1991}. 
In Sec. \ref{model},  we present our model for $\delta-\delta^{\prime}$ mirrors transparent at high frequencies and obtain the corresponding scattering coefficients. In Sec. \ref{Casimir-force} we compute the Casimir force and compare our results with those found in the literature. The final remarks are presented in Sec. \ref{final-remarks}.

\section{Generic semitransparent cavity}
\label{general}

Let us consider a general cavity formed by two mirrors, one of them labeled by the index $j=1$ and the other
by the index $j=2$, having scattering coefficients given by $r_{\pm}^{(j)}(\omega)$ and $s_{\pm}^{(j)}(\omega)$.
These scattering coefficients are the elements of the scattering matrix \cite{Jaekel-Reynaud-1991}:
\begin{eqnarray}
S^{(j)}(\omega)=\left(\begin{array}{cc}
s_{+}^{(j)}(\omega) & r_{+}^{(j)}(\omega)\mathrm{e}^{-2i\omega x_{j}}\\
r_{-}^{(j)}(\omega)\mathrm{e}^{2i\omega x_{j}} & s_{-}^{(j)}(\omega)
\end{array}\right),
\end{eqnarray}
where $x_{1} = 0$ and $x_{2} = q>0$ are the locations of the mirrors.
We request that the elements of the scattering matrix $S^{(j)}(\omega)$ obey the following conditions: 
\begin{enumerate}
\item Since the field is real  \cite{Jaekel-Reynaud-1991}:
\begin{eqnarray}
r_{\pm}^{(j)}(-\omega)&=&{r_{\pm}^{(j)}}^{*}(\omega),
\label{condicao-1-1}
\\
s_{\pm}^{(j)}(-\omega)&=&{s_{\pm}^{(j)}}^{*}(\omega). 	
\label{condicao-1-2}
\end{eqnarray}
\item Provided that the scattering matrix for each mirror is unitary  \cite{Jaekel-Reynaud-1991}:  
\begin{eqnarray}
|s_{\pm}^{(j)}(\omega)|^{2}+|r_{\pm}^{(j)}(\omega)|^{2}=1,
\label{condicao-2-1}
\\
s_{\pm}^{(j)}(\omega){r_{\mp}^{(j)}}^{*}(\omega)+r_{\pm}^{(j)}(\omega){s_{\mp}^{(j)}}^{*}(\omega)=0. 
\label{condicao-2-2}
\end{eqnarray}
\item Once the scattering is causal, what means that $s_{\pm}^{(j)}(\omega)$ and $r_{\pm}^{(j)}(\omega)$ vanish in the time domain for $t<0$  \cite{Jaekel-Reynaud-1991,Nussenzveig-1972}:
\begin{equation}
\begin{rcases}
  s_{\pm}^{(j)}(\omega) \\
  r_{\pm}^{(j)}(\omega) 
\end{rcases}
\;\text{are analytic for}\;\mathrm{Im}(\omega)>0.
\label{condicao-3}
\end{equation}
\item Requiring that the mirrors are transparent at high frequencies  \cite{Casimir-1948, Moore-1970, Jaekel-Reynaud-1991}:
\begin{equation}
\begin{rcases}
  |s_{\pm}^{(j)}(\omega)|=1 \\
  |r_{\pm}^{(j)}(\omega)|=0 
\end{rcases}
\;\text{for}\;\omega\rightarrow\infty.
\label{condicao-4}
\end{equation}
\end{enumerate}

The coefficients $r_{\pm}^{(j)}(\omega)$ are used in the formula for the Casimir 
force between two partially reflecting mirrors in $1+1$ dimensions, which can be computed as the difference between the outer and inner radiation pressures upon the mirrors, namely 
\begin{equation}
F_{\mathrm{L}}=T_{\mathrm{L}}-T_{\mathrm{cav}},\qquad F_{\mathrm{R}}=T_{\mathrm{cav}}-T_{\mathrm{R}},
\end{equation}
where $F_{\mathrm{L}}$ ($F_{\mathrm{R}}$) is the force exerted on
the mirror in the left (right), $T_{\mathrm{L}}$ ($T_{\mathrm{R}}$)
is the energy density in the left (right) of the cavity, and $T_{\mathrm{cav}}$
is the energy density within the cavity. For any partially reflecting
mirrors satisfying the aforementioned reality, unitarity, causality and transparency
conditions, the Casimir force is given by \cite{Jaekel-Reynaud-1991}
\begin{equation}
F=2\mathrm{Re}\int_{0}^{\infty}\frac{\mathrm{d}\omega}{2\pi}\frac{\omega r_{+}^{(1)}(\omega)r_{\mathrm{-}}^{(2)}(\omega)}{\mathrm{e}^{-2i\omega q}-r_{\mathrm{+}}^{(1)}(\omega)r_{\mathrm{-}}^{(2)}(\omega)},
\label{eq:aa}
\end{equation}
where $F=F_{\mathrm{R}}=-F_{\mathrm{L}}$,
assuming that the condition (\ref{condicao-4}) is satisfied so that 
\begin{equation}
\omega r_{+}^{(1)}(\omega)r_{\mathrm{-}}^{(2)}(\omega)\rightarrow0 \quad\text{for}\quad \omega\rightarrow\infty.
\label{convergencia}
\end{equation}
The integrand of Eq. (\ref{eq:aa}) vanishes for $|\omega|\rightarrow\infty$ and has no poles for $\mathrm{Im}(\omega)\ge0$ since $\left|r(\omega)\right|<1$ in this region. Therefore, using the Cauchy theorem, the integration path can be changed from the real to the imaginary axis \cite{Jaekel-Reynaud-1991}. It is preferable to work in the imaginary axis because the integrand of Eq. (\ref{eq:aa}) presents a simplified profile on it. Thereby, Eq. (\ref{eq:aa}) is replaced by 
\begin{eqnarray}
F & = & \int_{0}^{\infty}\mathcal{F}(\omega)\mathrm{d}\omega,
\label{Forca}
\\
\mathcal{F}(\omega) & = & -\frac{\omega}{\pi}\left[\frac{r_{\mathrm{+}}^{(1)}(i\omega)r_{\mathrm{-}}^{(2)}(i\omega)}{\mathrm{e}^{2\omega q}-r_{+}^{(1)}(i\omega)r_{\mathrm{-}}^{(2)}(i\omega)}\right].
\label{eq:esp-forca}
\end{eqnarray}
where the integrand is a function that oscillates less than the integrand in Eq. (\ref{eq:aa}). 

Next we propose a model for $\delta-\delta^{\prime}$ mirrors transparent at high frequencies and,
in Sec. \ref{Casimir-force}, we investigate the correspondent Casimir force.
%
\section{Modified $\delta-\delta^{\prime}$ model}
\label{model}

As mentioned in Sec. \ref{introduction}, the $\delta-\delta^{\prime}$ mirror described by (\ref{U-Munoz-0}) presents a problematic behavior at high frequencies [see Eq. (\ref{eq:hhh})]. 
In order to achieve full transparency at high frequencies, providing a more realistic description for $\delta-\delta^{\prime}$ mirrors, we start proposing the following Lagrangian density
\begin{eqnarray}
{\cal L}&=&{\cal L}_{0}-[\mu\delta(x)+\lambda\delta^{\prime}(x)]\Phi^{2}(t,x)
\cr\cr
&&-\delta^{\prime}(x)\sum_{n=1}^{\infty}\rho_{n}(\Lambda)[\partial_{t}^{n}\Phi(t,x)]^{2},
\label{nossa-proposta-1-espelho}
\end{eqnarray}
where 
\begin{equation}
\rho_{n}(\Lambda)=\lambda\int_{0}^{\Lambda}\frac{\mathrm{d}u}{2\pi}\frac{4\beta}{\beta^{2}+u^{2}}\frac{(-1)^{n}u^{2n}}{(2n)!},
\label{nossa-escolha}
\end{equation}
If 
$\Lambda=0$ or $\beta=0$ one recovers the model (\ref{U-Munoz-0}) found
in the literature \cite{Castaneda-2015}.
This Lagrangian yields the field equation 
\begin{eqnarray}
\left[\partial_{t}^{2}-\partial_{x}^{2}+2\mu\delta(x)+2\lambda\delta^{\prime}(x)\right]\Phi(t,x)
\nonumber\\  
+\delta^{\prime}(x)\sum_{n=1}^{\infty}2(-1)^{n}\rho_{n}(\Lambda)\partial_{t}^{2n}\Phi(t,x)&=&0.
\label{MovEq-time}
\end{eqnarray}
In the Fourier domain, Eq. (\ref{MovEq-time}) reads
\begin{eqnarray}
\left[-\omega^{2}-\partial_{x}^{2}+2\mu\delta(x)+2\lambda\delta^{\prime}(x)\right]\phi(\omega,x)\nonumber\\
+2\delta^{\prime}(x)\sum_{n=1}^{\infty}\rho_{n}(\Lambda)\omega^{2n}\phi(\omega,x)&=&0.
\label{MovEq-n}
\end{eqnarray}
Since the functions $\rho_n(\Lambda)$ are integrable on the interval $[0, \Lambda]$, and 
given the convergence
\begin{equation}
\sum_{n=1}^{\infty}\frac{(-1)^{n}u^{2n}}{(2n)!}\omega^{2n}=-1+\cos(u\omega)
\end{equation}
on that interval, one can write \cite{Gradshteyn-livro} 
\begin{eqnarray}
\sum_{n=1}^{\infty}\rho_{n}(\Lambda)\omega^{2n}&=&\lambda\int_{0}^{\Lambda}\frac{\mathrm{d}u}{2\pi}\frac{4\beta}{\beta^{2}+u^{2}}
\sum_{n=1}^{\infty}\frac{(-1)^{n}u^{2n}}{(2n)!}\omega^{2n}\nonumber\\
&=&\lambda\int_{0}^{\Lambda}\frac{\mathrm{d}u}{2\pi}\frac{4\beta}{\beta^{2}+u^{2}}
[-1+\cos(u\omega)].
\end{eqnarray}

The $\delta-\delta^{\prime}$ model we are interested in is that obtained from (\ref{nossa-proposta-1-espelho}) in the limit $\Lambda\rightarrow\infty$.
Taking into account that
\begin{equation}
\lim_{\Lambda\rightarrow\infty}\int_{0}^{\Lambda}\frac{\mathrm{d}u}{2\pi}\frac{4\beta}{\beta^{2}+u^{2}}\cos(u\omega)=\exp(-\beta\left|\omega\right|),
\end{equation}
the field equation can be written as
\begin{equation}
[-\partial_{x}^{2}+2\mu\delta(x)+2\lambda\exp(-\beta\left|\omega\right|)\delta^{\prime}(x)]\phi(\omega,x)=\omega^{2}\phi(\omega,x).
\label{MovEq}
\end{equation}
Next, considering the model described by the field equation (\ref{MovEq}), we will find the corresponding matching conditions and the reflection and transmission coefficients.

The field and its spatial derivative are not considered to be continuous at $x=0$, \textit{a priori}. Therefore, let us consider the following properties of the $\delta$ and $\delta^{\prime}$ functions (see, for instance, Refs. \cite{Kurasov-1996-1,Gadella-2009-1}): 
\begin{eqnarray}
\delta(x)\phi(\omega,x) & = & \frac{\phi(\omega,0^{+})+\phi(\omega,0^{-})}{2}\delta(x),\label{AA1}\\
\delta^{\prime}(x)\phi(\omega,x) & = & \frac{\phi(\omega,0^{+})+\phi(\omega,0^{-})}{2}\delta^{\prime}(x)\nonumber \\
 &  & -\frac{\phi^{\prime}(\omega,0^{+})+\phi^{\prime}(\omega,0^{-})}{2}\delta(x).\qquad\label{AA2}
\end{eqnarray}
Substituting Eqs. (\ref{AA1}) and (\ref{AA2}) in (\ref{MovEq}) and integrating
across $x=0$, we obtain 
\begin{eqnarray}
-\phi^{\prime}(\omega,0^{+})+\phi^{\prime}(\omega,0^{-})+\mu\left[\phi(\omega,0^{+})+\phi(\omega,0^{-})\right]\nonumber \\
-\lambda\exp(-\beta\left|\omega\right|)\left[\phi^{\prime}(\omega,0^{+})+\phi^{\prime}(\omega,0^{-})\right] & = & 0.\nonumber \\
\label{AA0}
\end{eqnarray}
Now, integrating Eq. (\ref{MovEq}) twice, the first one from $-L<0$ to $x$
(see, for instance, Ref. \cite{Kurasov-1993}) resulting in
\begin{eqnarray}
-\phi^{\prime}(\omega,x)+\phi^{\prime}(\omega,-L)+\mu\left[\phi(\omega,0^{+})+\phi(\omega,0^{-})\right]\Theta(x)\nonumber \\
-\lambda\exp(-\beta\left|\omega\right|)\left[\phi^{\prime}(\omega,0^{+})+\phi^{\prime}(\omega,0^{-})\right]\Theta(x)\nonumber \\
+\lambda\exp(-\beta\left|\omega\right|)\left[\phi(\omega,0^{+})+\phi(\omega,0^{-})\right]\delta(x)\nonumber \\
=\omega^{2}\int_{-L}^{x}\phi(\omega,x)\;\mathrm{d}x,\nonumber \\
\label{AA-intermed}
\end{eqnarray}
and integrating across $x=0$, we obtain 
\begin{eqnarray}
-\phi(\omega,0^{+})+\phi(\omega,0^{-})\nonumber \\
+\lambda\exp(-\beta\left|\omega\right|)[\phi(\omega,0^{+})+\phi(\omega,0^{-})]&=&0.
\label{AA3}
\end{eqnarray}

Combining Eqs. (\ref{AA0}) and (\ref{AA3}) we get the matching conditions (see, for instance, Refs. \cite{Gadella-2009-1,Kurasov-1996-1,Castaneda-2015,JD-2016}),
\begin{equation}
\phi(\omega,0^{+})=\frac{1+\lambda\exp(-\beta\left|\omega\right|)}{1-\lambda\exp(-\beta\left|\omega\right|)}\phi(\omega,0^{-}),\label{eq:MC1}
\end{equation}
\begin{eqnarray}
\phi^{\prime}(\omega,0^{+})&=&\frac{1-\lambda\exp(-\beta\left|\omega\right|)}{1+\lambda\exp(-\beta\left|\omega\right|)}\phi^{\prime}(\omega,0^{-})\nonumber\\
&&+\frac{2\mu}{1-\lambda^2\exp(-2\beta\left|\omega\right|)}\phi(\omega,0^{-}).\label{eq:MC2}
\end{eqnarray}

The solution of Eq. (\ref{MovEq-time}) can be written as
\begin{equation}
\Phi(t,x) = \sum_{j=L,R}\int_{0}^{\infty}\mathrm{d}\omega[a_{j}(\omega)\Psi_{j}(\omega,x)\mathrm{e}^{-i\omega t}+H.c.],
\end{equation}
where
\begin{eqnarray}
\Psi_{R}(\omega,x)&=&\frac{1}{\sqrt{4\pi\omega}}\left\{ \Theta(x)\left[r_{+}(\omega)\mathrm{e}^{i\omega x}+\mathrm{e}^{-i\omega x}\right]
\right.
\cr\cr
&&\left.+\Theta(-x)s_{-}(\omega)\mathrm{e}^{-i\omega x}\right\}, 
\label{psi-R}
\end{eqnarray}
\begin{eqnarray}
\Psi_{L}(\omega,x)&=&\frac{1}{\sqrt{4\pi\omega}}\left\{ \Theta(-x)\left[r_{-}(\omega)\mathrm{e}^{-i\omega x}+\mathrm{e}^{i\omega x}\right]
\right.
\cr\cr
&&\left.+\Theta(x)s_{+}(\omega)\mathrm{e}^{i\omega x}\right\}
\label{psi-L}
\end{eqnarray}
are the right- and left-incident solutions (see, for instance, Ref. \cite{Barton-I-II-1995}), $H.c.$ is the Hermitian conjugate of the previous term, and $a_j(\omega)$ ($j=L,R$) are the annihilation operators, obeying $[a_i(\omega),a_j^{\dagger}(\omega^{\prime})]=\delta(\omega-\omega^{\prime})\delta_{ij}$.
Eqs. (\ref{eq:MC1}) and (\ref{eq:MC2}), together with Eqs. (\ref{psi-R}) and (\ref{psi-L}), lead to the coefficients
\begin{equation}
r_{\mathrm{\pm}}(\omega)=\frac{\pm2\omega \lambda\exp(-\beta\left|\omega\right|)-i\mu}{\omega[1+\lambda^2\exp(-2\beta\left|\omega\right|)]+i\mu},
\label{eq:refle-1}
\end{equation}
\begin{equation}
\\ s_{\pm}(\omega)=\frac{\omega[1-\lambda^2\exp(-2\beta\left|\omega\right|)]}{\omega[1+\lambda^2\exp(-2\beta\left|\omega\right|)]+i\mu},
\label{eq:refle}
\end{equation}
Now, the conditions in Eq. (\ref{condicao-4}) are satisfied, so that for sufficiently high frequencies the mirror described 
by Eq. (\ref{MovEq}) does not act as an obstacle to the field, as expected for realistic models. 
In addition, the scattering coefficients (\ref{eq:refle-1}) and (\ref{eq:refle}) obey the conditions given in Eqs. (\ref{condicao-1-1})$-$(\ref{condicao-3}). At this point, we remark that 
the function $\rho_{n}(\Lambda)$ given in Eq. (\ref{nossa-escolha}) belongs to a class of functions which, following a way similar
to that we used from Eq. (\ref{nossa-escolha}) to (\ref{MovEq}), can result in
\begin{equation}
r_{\mathrm{\pm}}(\omega)=\frac{\pm2\omega \lambda f(\left|\omega\right|))-i\mu}{\omega[1+\lambda^2f(\left|\omega\right|)^2]+i\mu},
\label{eq:refle-1-general}
\end{equation}
\begin{equation}
\\ s_{\pm}(\omega)=\frac{\omega[1-\lambda^2f(\left|\omega\right|)^2]}{\omega[1+\lambda^2f(\left|\omega\right|)^2]+i\mu},
\label{eq:refle-general}
\end{equation}
with $f(|\omega|)$ given so that (\ref{eq:refle-1-general}) and (\ref{eq:refle-general}) obey the conditions (\ref{condicao-1-1})$-$(\ref{condicao-4}). For instance, $\rho_{n}(\Lambda)$ could be chosen so that 
$f(\left|\omega\right|)=\exp(-\beta\left|\omega\right|^2)$, among other infinite possibilities.
\begin{figure}[t] 
\centering
\includegraphics[width=0.87\columnwidth]{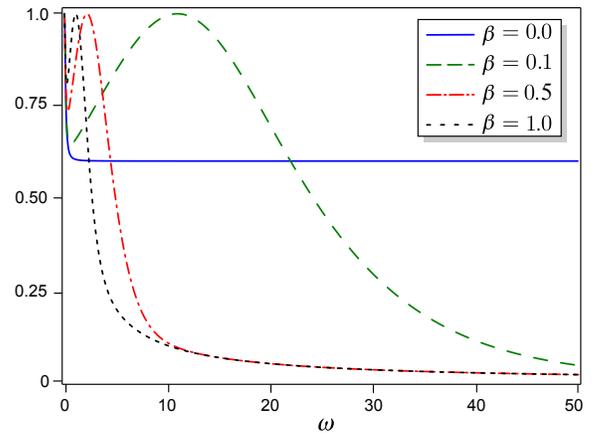} 
\caption{$|r_{+}(\omega)|$ (vertical axis), for $\mu = 1$, $\lambda = 3$, and several values of $\beta$. The solid line ($\beta = 0$), shows the reflection coefficient correspondent to the $\delta-\delta^{\prime}$ model given by Eq. (\ref{U-Munoz-0}). The dashed
($\beta=0.1$), dot-dashed ($\beta=5.0$), and space-dashed ($\beta=1.0$) lines correspond to $|r_{+}(\omega)|$ for the modified $\delta-\delta^{\prime}$ model given by Eq. (\ref{nossa-proposta-1-espelho}). The behavior of $|r_{-}(\omega)|$ is similar.}
\label{modulo-r}
\end{figure}

In Fig. \ref{modulo-r}, we show $|r_{+}(\omega)|$, given by Eq. (\ref{eq:refle-1}), for several values of the parameter $\beta$.
When $\beta \neq 0$, $|r_{+}(\omega)|$ vanishes as $\omega\rightarrow\infty$ (see the dashed, dot-dashed, and space-dashed lines in Fig. \ref{modulo-r}), whereas for $\beta = 0$, which recovers the $\delta-\delta^{\prime}$ model given by Eq. (\ref{U-Munoz-0}), $|r_{+}(\omega)|$ becomes a non-zero constant (see the solid line in Fig. \ref{modulo-r}), as shown in Eq. (\ref{eq:hhh}).

In the case of two mirrors, a Lagrangian which extends (\ref{nossa-proposta-1-espelho})  is
\begin{eqnarray}
{\cal L} & = & \mathcal{L}_{0}-\sum_{j}\left[{\mu}_{j}\delta(x-x_{j})+\lambda_j\delta^{\prime}(x-x_{j})\right]\Phi^{2}(t,x)\nonumber \\
 &  & -\sum_{j}\delta^{\prime}(x-x_{j})\sum_{n=1}^{\infty}{\rho_{n}}_{j}(\Lambda)[\partial_{t}^{n}\Phi(t,x)]^{2},\quad
\label{two-mirrors}
\end{eqnarray}
where being $j=1,2$ indicates each mirror, and $x_{1} = 0$ and $x_{2} = q$ are the locations of the mirrors. The matching conditions for each mirror are still given by Eqs. (\ref{eq:MC1}) and (\ref{eq:MC2}), but with $\mu$, $\lambda$ and $\beta$ labeled with $j$, and $0^{\pm}$ replaced by $x_j^{\pm}$. Therefore, analogously to the case of a single mirror, in the limit $\Lambda\rightarrow\infty$ we find that the coefficients associated to (\ref{two-mirrors}) are
\begin{equation}
r_{\mathrm{\pm}}^{(j)}(\omega)=\frac{\pm2\omega\lambda_{j}\exp(-\beta_{j}|\omega|)-i{\mu}_{j}}{\omega[1+\lambda_{j}^2\exp(-2\beta_{j}|\omega|)]+i{\mu}_{j}},
\label{Reflex}
\end{equation}
\begin{equation}
s_{\pm}^{(j)}(\omega)=\frac{\omega[1-\lambda_{j}^2\exp(-2\beta_{j}|\omega|)]}{\omega[1+\lambda_{j}^2\exp(-2\beta_{j}|\omega|)]+i{\mu}_{j}}.
\label{Transm}
\end{equation}

Next, with the aid of Eq. (\ref{Reflex}), we investigate the Casimir force between the two mirrors described by Eq. (\ref{two-mirrors}).

\section{Casimir force}
\label{Casimir-force}
\begin{figure}[t]
\begin{centering}
\includegraphics[width=0.999\columnwidth]{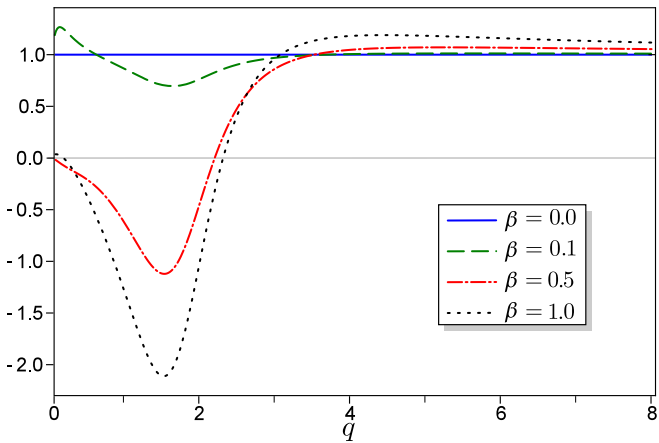} 
\par\end{centering}
\caption{$F|_{\beta \neq 0}$/$F|_{\beta=0}$ as a function of $q$,
considering the parameters $\mu_{1}= 1$, $\mu_{2} = 3$, $\lambda_1 = 3$, $\lambda_2=-2$ and $\beta_1 = \beta_2 = \beta$.}
\label{taxa}
\end{figure}
To obtain the Casimir force between the two $\delta-\delta^{\prime}$ mirrors modeled by the Lagrangian density (\ref{two-mirrors}), we substitute Eq. (\ref{Reflex}) into (\ref{eq:esp-forca}) and get
\begin{figure*}[t]
\centering
\subfloat[$\beta=0$.]{
\includegraphics[width=0.98\columnwidth]{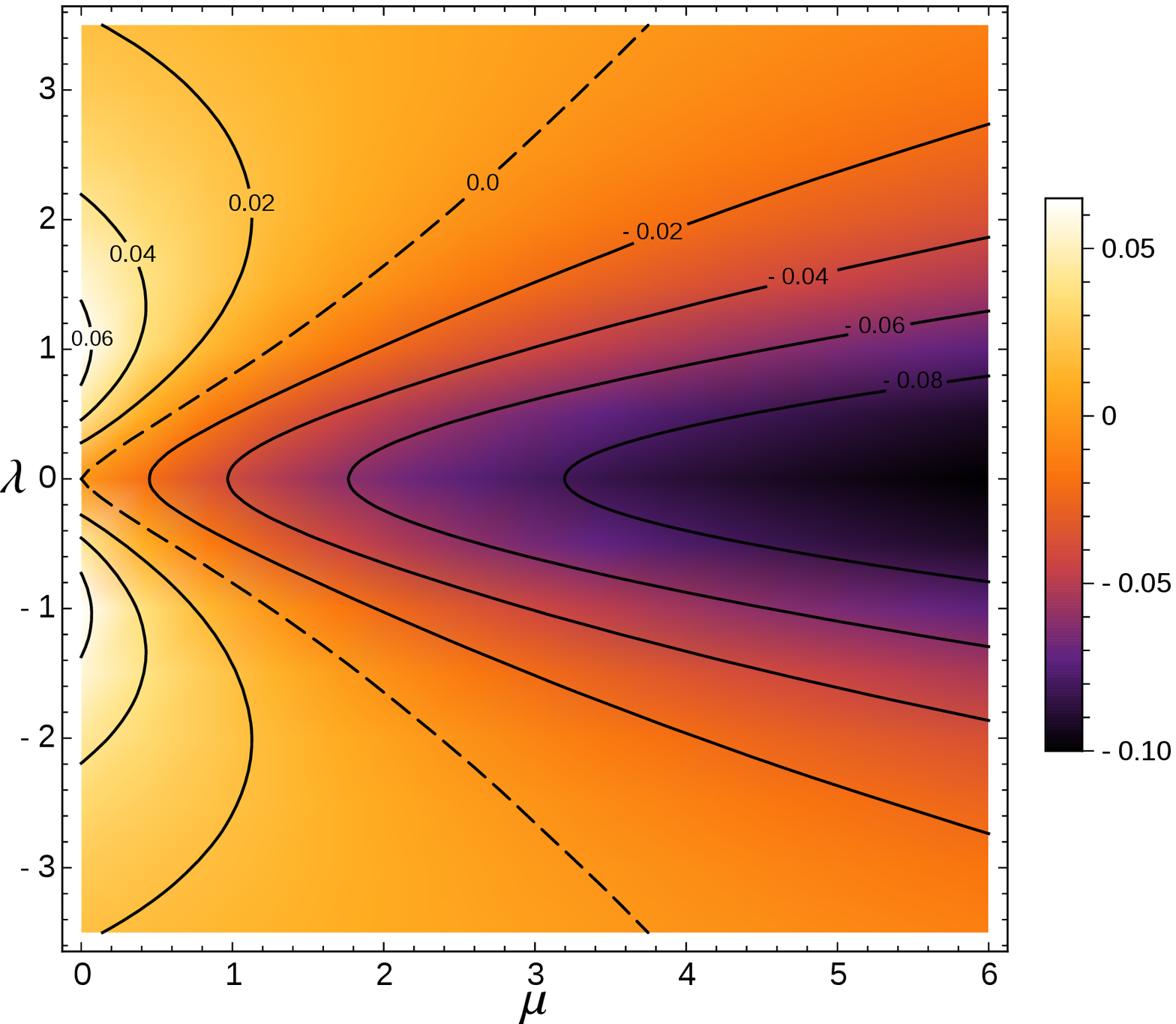}
\label{modulo-plano-mu-lambda-1}
}
\quad
\subfloat[$\beta=1$.]{
\includegraphics[width=0.98\columnwidth]{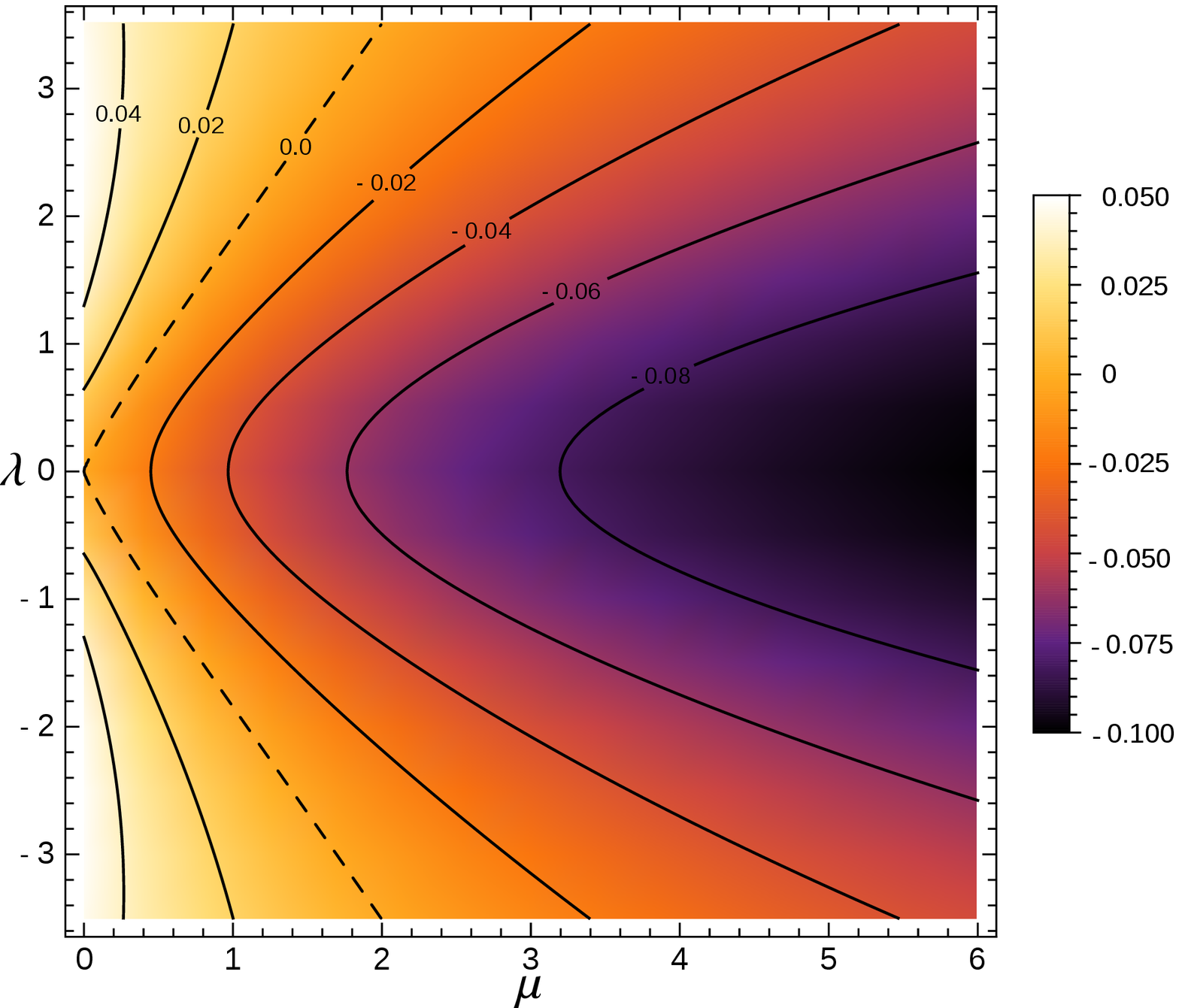}
\label{modulo-plano-mu-lambda-2}
}
\caption{Casimir force as function of $\mu=\mu_1=\mu_2$ and $\lambda=\lambda_1=\lambda_2$, considering $q = 1$ and two different values for $\beta=\beta_1=\beta_2$. Fig. \ref{modulo-plano-mu-lambda-1} shows $F$ with $\beta=0$ [recovering the results for the model given by Eq. (\ref{U-Munoz-0})], and Fig. \ref{modulo-plano-mu-lambda-2} exhibits $F$ with $\beta=1$. The solid lines are the level curves, and the dashed lines show where the Casimir force is null.}
\label{modulo-plano-mu-lambda}
\end{figure*}
\begin{equation}
\mathcal{F}(\omega)=-\frac{1}{\pi}\frac{\omega}{1-\mathrm{e}^{2\omega q}\mathcal{B}_{1}(\omega)\mathcal{B}_{2}(\omega)},
\label{eq:forca-delta-delta-p-1}
\end{equation}
where
\begin{equation}
\mathcal{B}_{j}(\omega)=\frac{\omega[1+\lambda_{j}^2\exp(-2\beta_{j}|\omega|)]
+{\mu}_{j}}{(-1)^{j}2\omega[\lambda_{j}\exp(-\beta_{j}|\omega|)]+{\mu}_{j}}.
\label{B-j}
\end{equation}
The Casimir force is obtained after integrating $\mathcal{F}(\omega)$ as shown in Eq. (\ref{Forca}).
\begin{figure}[t] 
\centering
\includegraphics[width=0.98\columnwidth]{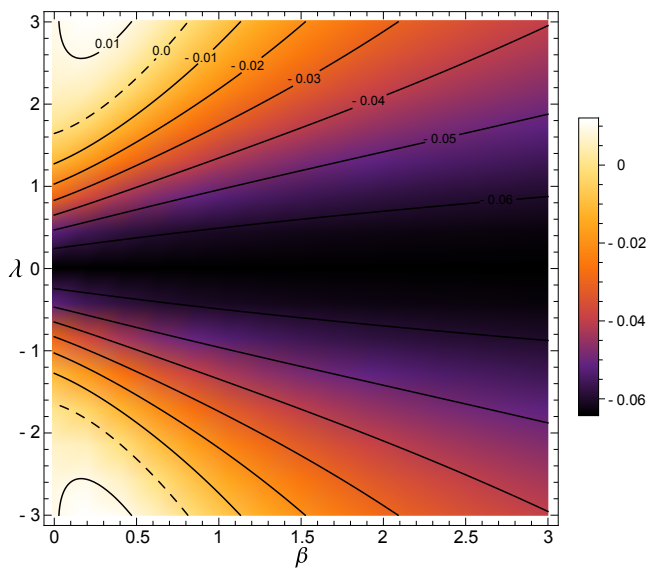} 
\caption{Casimir force as function of $\lambda$ and $\beta$, considering $\lambda_1=\lambda_2=\lambda$, $\beta_1=\beta_2=\beta$, $\mu_1=\mu_2=2$, and $q=1$.}
\label{fig:beta-lambda}
\end{figure}

Due to the exponential term $\mathrm{e}^{2\omega q}$ in Eq. (\ref{eq:forca-delta-delta-p-1}),
one can find a certain frequency $\omega_q$ for which
$\mathcal{F}(\omega>\omega_q)\approx 0$, so that one has, in a good approximation, 
\begin{eqnarray}
F\approx \int_{0}^{\omega_q}\mathcal{F}(\omega)\mathrm{d}\omega,
\label{troca}
\end{eqnarray}
where $\omega_q$ decreases as $q$ increases. 
On the other hand, from Eq. (\ref{B-j}), we can see that as $\omega\rightarrow 0$, $\mathcal{B}_{j}|_{\beta\neq0}\rightarrow \mathcal{B}_{j}|_{\beta=0}$.
Therefore, as $q$ increases, the upper limit of the integral in Eq. (\ref{troca}) becomes smaller and, thus,
$F|_{\beta\neq0}\rightarrow F|_{\beta=0}$.
This can be visualized in Fig. \ref{taxa}, where we show the ratio between the Casimir force for $\beta_j\neq 0$ and the Casimir force for $\beta_j=0$ as a function of the distance $q$ between two different mirrors.
Notice that $F|_{\beta\neq0}/F|_{\beta=0}\rightarrow 1$ as $q$ increases. As $q$ decreases, one may have 
discrepancies between the Casimir force provided by both models.

In Fig. \ref{modulo-plano-mu-lambda}, we exhibit the Casimir force as function
of $\mu_1=\mu_2=\mu$ and $\lambda_1=\lambda_2=\lambda$, for a fixed distance $q=1$ and two different values for $\beta_1=\beta_2=\beta$: Fig. \ref{modulo-plano-mu-lambda-1} for $\beta=0$ (recovering the case found in the literature \cite{Castaneda-2015}), and Fig. \ref{modulo-plano-mu-lambda-2} for $\beta=1$. 
By comparing Figs. \ref{modulo-plano-mu-lambda-1} and \ref{modulo-plano-mu-lambda-2}, we see that the $\beta$ parameter
can increase or decrease the magnitude of the Casimir force ($|F|$), and can also change its sign, depending on the value of $\beta$, in comparison with the case where $\beta=0$.
For instance, considering $\mu=3$ and $\lambda= 2$ in Figs. \ref{modulo-plano-mu-lambda-1} and \ref{modulo-plano-mu-lambda-2}, we see that 
$|F|$ when $\beta=0$ is smaller than that for $\beta=1$, and in both cases the force is attractive. 
But, for $\mu=1$ and $\lambda= 2$,
$|F|$ when $\beta=0$ is greater than that for $\beta=1$, and in both cases the force is repulsive.

In Fig. \ref{fig:beta-lambda}, we exhibit the Casimir force as function
of $\beta_1=\beta_2=\beta$ and $\lambda_1=\lambda_2=\lambda$, for a fixed distance $q=1$ and $\mu_1=\mu_2=2$.
We can see that the Casimir force tends to an attractive behavior if $\beta$ increases. 
In fact, if we take the limit $\beta\rightarrow\infty$ in Eqs. (\ref{eq:refle-1}) and (\ref{eq:refle}), we
recover the coefficients in Eq. (\ref{r-s-delta}), which describe partially reflecting mirrors associated with a
purely attractive Casimir force.

\section{Final remarks\label{final-remarks}}

Although partially reflecting mirrors simulated by Dirac $\delta$
point interactions [Eq. (\ref{lagrangeana-U})] are naturally transparent at high frequencies [Eq. (\ref{barton-transparencia})],
the same is not valid [Eq. (\ref{eq:hhh})] for mirrors simulated by the Dirac $\delta-\delta^{\prime}$
point interactions modeled by (\ref{U-Munoz-0}). 
Moreover, for $\mu=0$ the model (\ref{U-Munoz-0}) gives scattering coefficients independent of the frequency [Eq. (\ref{eq:rr-1-2})].
Since, for sufficiently high frequencies, a realistic mirror is full transparent \cite{Casimir-1948, Moore-1970,Jaekel-Reynaud-1991},
and mirrors that violate this property can, in some situations, lead to unphysical predictions \cite{Haro-Elizalde-PRL-2006},
in order to provide more realistic features for $\delta-\delta^{\prime}$ models,
we proposed the lagrangian given by Eq. (\ref{nossa-proposta-1-espelho}).
This model (\ref{nossa-proposta-1-espelho}), in the limit $\Lambda\rightarrow\infty$,
provides the coefficients (\ref{eq:refle-1}) and (\ref{eq:refle}) leading to full transparency in the limit of high frequencies 
[Eq. (\ref{condicao-4})] (see also Fig. \ref{modulo-r}).
Considering a cavity formed by two of these $\delta-\delta^{\prime}$ mirrors [Eq. (\ref{two-mirrors})], 
we investigated the behavior of the Casimir force, comparing our results
with the Casimir force between partially reflecting mirrors modeled by (\ref{U-Munoz-0}) \cite{Castaneda-2015}, this can be visualized in the $\mu\lambda$-plane shown in Fig. \ref{modulo-plano-mu-lambda}. In this figure, we showed that the $\beta$-terms in the Lagrangian (\ref{two-mirrors}) can change the sign of the Casimir force, and can increase or decrease the magnitude of the Casimir force, 
in comparison with the case where $\beta=0$.
As the distance between the mirrors increases, the Casimir force for ${\beta\neq0}$ tends to the Casimir force for ${\beta=0}$.
Conversely, for the distance decreasing, one may have discrepancies between the Casimir force provided by both models
(see Fig. \ref{taxa}).

\begin{acknowledgments}
We thank D. C. Pedrelli, F. S. S. Rosa, V. S. S. Alves, C. Farina, A. L. C. Rego and C. L. Benone for valuable discussions. This work was partially supported by CNPq and CAPES\textemdash Brazil. 
\end{acknowledgments}

\end{document}